\pdfoutput=1

\documentclass{jors}

\begin{document}

{\bf Software paper for submission to the Journal of Open Research Software} \\

\rule{\textwidth}{1pt}

\section*{(1) Overview}

\vspace{0.5cm}

\section*{Title}

FluidSim: modular, object-oriented Python package for high-performance CFD
simulations

\section*{Paper Authors}

1. MOHANAN, Ashwin Vishnu$^a$ \\
2. BONAMY, Cyrille$^b$ \\
3. CALPE LINARES, Miguel$^b$ \\
4. AUGIER, Pierre$^b$

\smallskip

$^a$ Linn\'e Flow Centre, Department of Mechanics, KTH, 10044 Stockholm, Sweden.\\
$^b$ Univ. Grenoble Alpes, CNRS, Grenoble INP\footnote{Institute of Engineering
Univ. Grenoble Alpes}, LEGI, 38000 Grenoble, France.

\section*{Paper Author Roles and Affiliations}
1. Ph.D. student, Linn\'e Flow Centre, KTH Royal Institute of Technology,
Sweden; \\
2. Research Engineer, LEGI, Universit\'e Grenoble Alpes, CNRS, France; \\
3. Ph.D. student, LEGI, Universit\'e Grenoble Alpes, CNRS, France
4. Researcher, LEGI, Universit\'e Grenoble Alpes, CNRS, France

\section*{Abstract}


The Python package \fluidpack{sim} is introduced in this article as an extensible
framework for Computational Fluid Mechanics (CFD) solvers.
It is developed as a part of \href{https://fluiddyn.readthedocs.io}{FluidDyn}
project \citep{fluiddyn}, an effort to promote open-source and open-science
collaboration within fluid mechanics community and intended for both educational
as well as research purposes.
Solvers in \fluidpack{sim} are scalable, High-Performance Computing (HPC) codes
which are powered under the hood by the rich, scientific Python ecosystem and the
Application Programming Interfaces (API) provided by \fluidpack{dyn} and
\fluidpack{fft} packages \citep{fluidfft}.
The present article describes the design aspects of \fluidpack{sim}, viz.\ use
of Python as the main language; focus on the ease of use, reuse and maintenance
of the code without compromising performance.
The implementation details including optimization methods, modular organization
of features and object-oriented approach of using classes to implement solvers
are also briefly explained.
Currently, \fluidpack{sim} includes solvers for a variety of physical problems
using different numerical methods (including finite-difference methods).
However, this metapaper shall dwell only on the implementation and performance
of its pseudo-spectral solvers, in particular the two- and three-dimensional
Navier-Stokes solvers.
We investigate the performance and scalability of \fluidpack{sim} in a
state of the art HPC cluster.
Three similar pseudo-spectral CFD codes based on Python (Dedalus, SpectralDNS) and
Fortran (NS3D) are presented and qualitatively and quantitatively compared to
\fluidpack{sim}.
The source code is hosted at Bitbucket as a Mercurial repository
\href{https://bitbucket.org/fluiddyn/fluidsim}{bitbucket.org/fluiddyn/fluidsim}
and the documentation generated using Sphinx can be read online at
\href{https://fluidsim.readthedocs.io}{fluidsim.readthedocs.io}.

\section*{Keywords}

Python;\ CFD;\ HPC;\ MPI;\ modular;\ object-oriented;\ tested;\ documented;\
open-source

\section*{Introduction}



Designed as a specialized package of the
\href{https://fluiddyn.readthedocs.io}{FluidDyn} project for computational fluid
mechanics (CFD), \fluidpack{sim} is a comprehensive solution to address the needs
of a fluid mechanics student and researcher alike --- by providing scalable high
performance solvers, on-the-fly postprocessing, and plotting functionalities under
one umbrella.  In the open-science paradigm, scientists will be both users and
developers of the tools at the same time.
An advantage of \fluidpack{sim} is that, most of the users just have to read and
write Python code.  \fluidpack{sim} ensures that all critical modules, classes and
functions are well documented --- both as inline comments and as standalone
documentation, complete with examples and tutorials.  For these reasons
\fluidpack{sim} can become a true collaborative code and has the potential to
replace some in-house pseudo-spectral codes written in more conventional
languages.

\subsubsection*{Balance between runtime efficiency and cost of development}
In today's world where clusters are moving from petascale to exascale
performance, computing power is aplenty. In such a scenario, it becomes
apparent that \emph{man-hours are more expensive than computing time}. In other
words, the cost of development outweighs the cost of computing time. Therefore,
developers should be willing to make small sacrifices in efficiency, to improve
development time, code maintainability and clarity in general.

For the above reasons, majority of \fluidpack{sim}'s code-base, in terms of
line of code, is written using pure Python syntax. However, this is done
without compromising performance, by making use of libraries such as \Numpy,
and optimized compilers such as \pack{Cython} and \pack{Pythran}.

\Numpy functions and data types are sufficient for applications such as
initialization and postprocessing operations, since these functions are used
sparingly. Computationally intensive tasks such as time-stepping and linear
algebra operators which are used in every single iteration must be offloaded to
compiled extensions.
This optimization strategy can be considered as the computational equivalent of
the \href{https://en.wikipedia.org/wiki/Pareto_principle}{Pareto principle}, also
known as the 80/20 rule\footnote{See \citet{behnel_cython2011},
\href{https://wiki.haskell.org/Why_Haskell_matters}{%
wiki.haskell.org/Why\_Haskell\_matters}}.
The goal is to optimize such that ``80 percent of the runtime is spent in 20
percent of the source code'' \cite[]{meyers2012effective}.
Here, \pack{Cython} \citep{behnel_cython2011} and \pack{Pythran}
\citep{guelton2018pythran} compilers comes in handy.
%
An example on how we use \pack{Pythran} to reach similar performance than with
Fortran by writing only Python code is described in the companion paper on
\fluidpack{fft} \citep{fluidfft}.




The result of using such an approach is studied in the forthcoming sections by
measuring the performance of \fluidpack{sim}.
We will demonstrate that a very large percentage of the elapsed time is spent in
the execution of optimized compiled functions and thus that the ``Python cost'' is
negligible.

\subsubsection*{Target audiences}
\fluidpack{sim} is designed to cater to the needs of three kinds of audience.
\begin{itemize}
\item \emph{Users}, who run simulations with already available solvers. To do
this, one needs to have very basic skills in Python scripting.
\item \emph{Advanced users}, who may extend \fluidpack{sim} by developing
customized solvers for new problems by taking advantage of built-in operators and
time-stepping classes.  In practice, one can easily implement such solvers by
changing a few methods in base solver classes.  To do this, one needs to have
fairly good skills in Python and in particular object-oriented programming.
\item \emph{Developers}, who develop the base classes, in particular, the
operators and time stepping classes.  One may also sometime need to write compiled
extensions to improve runtime performance. To do this, desirable traits include
good knowledge in Python, \Numpy, \pack{Cython} and \pack{Pythran}.
\end{itemize}

This metapaper is intended as a short introduction to \fluidpack{sim} and its
implementation, written mainly from a user-perspective. Nevertheless, we also discuss how
\fluidpack{sim} can be customized and extended with minimal effort to promote code
reuse.
A more comprehensive and hands-on look at how to use \fluidpack{sim} can be found
in the tutorials\footnote{See
\href{https://fluidsim.readthedocs.io/en/latest/tutorials.html}{fluidsim.readthedocs.io/en/latest/tutorials.html}},
both from a user's and a developer's perspective.
In the latter half of the paper, we shall also inspect the performance of
\fluidpack{sim} in large computing clusters and compare \fluidpack{sim} with three
different pseudo-spectral CFD codes.

\section*{Implementation and architecture}


New features were added over the years to the package whenever demanded by
research interests, thus making the code very user-centric and
function-oriented.  This aspect of code development is termed as YAGNI, one of
the principles of agile programming software development method, which
emphasizes not to spent a lot of time developing a functionality, because most
likely \emph{you aren't gonna need it}.

Almost all functionalities in \fluidpack{sim} are implemented as classes and its
methods are designed to be modular and generic.  This means that the user is free
to use inheritance to modify certain parts to suit one's needs, and avoiding the
risk of breaking a functioning code.

\subsubsection*{Package organization}

\fluidpack{sim} is meant to serve as a framework for numerical solvers using
different methods. For the present version of \fluidpack{sim} there is support for
finite difference and pseudo-spectral methods. An example of a finite difference
solver is \codeinline{fluidsim.solvers.ad1d} which solves the 1D advection
equation. There are also solvers which do not rely on most of the base classes,
such as \codeinline{fluidsim.base.basilisk} which implements a 2D adaptive meshing
solver based on the CFD code \href{http://basilisk.fr/}{Basilisk}. The collection
of solvers using pseudo-spectral methods are more feature-rich in comparison.

The code is organized into the following sub-packages:

\begin{itemize}
\item \codeinline{fluidsim.base}: contains all base classes and a solver for the
trivial equation $\partial_t \mathbf{\hat{u}} = 0 $.
\item \codeinline{fluidsim.operators}: specialized linear algebra and numerical
method operators (e.g., divergence, curl, variable transformations, dealiasing).

\item \codeinline{fluidsim.solvers}: solvers and postprocessing modules for
problems such as 1D advection, 2D and 3D Navier-Stokes equations (incompressible
and under the Boussinesq approximation, with and without density stratification
and system rotation), one-layer shallow water and F\"oppl-von K\'arm\'an
equations.

\item \codeinline{fluidsim.util}: utilities to load and modify an existing
simulation, to test, and to benchmark a solver.
\end{itemize}

Subpackages \codeinline{base} and \codeinline{operators} form the backbone of this
package, and are not meant to be used by the user explicitly.
In practice, one can make an entirely new solver for a new problem using this
framework by simply writing one or two importable files containing three classes:
\begin{itemize}
\item an \codeinline{InfoSolver} class\footnote{Inheriting from the base class
\codeinline{fluidsim.base.solvers.info\_base.InfoSolverBase}.}, containing the
information on which classes will be used for the different tasks in the solver
(time stepping, state, operators, output, etc.).
\item a \codeinline{Simulation} class\footnote{Inheriting from the base class
\codeinline{fluidsim.base.solvers.base.SimulBase}.} describing the equations to be
solved.
\item a \codeinline{State} class\footnote{Inheriting from the base class
\codeinline{fluidsim.base.state.StateBase}.} defining all physical variables and
their spectral counterparts being solved (for example: $u_x$ and $u_y$) and
methods to compute one variable from another.
\end{itemize}

We now turn our attention to the simulation object which illustrates how to
access the code from the user's perspective.

\subsubsection*{The simulation object}

The simulation object is instantiated with necessary parameters just before
starting the time stepping.  A simple 2D Navier-Stokes simulation can be
launched using the following Python script:

\begin{minted}[fontsize=\footnotesize]{python}
 from fluidsim.solvers.ns2d.solver import Simul

 params = Simul.create_default_params()
 # Modify parameters as needed
 sim = Simul(params)
 sim.time_stepping.start()
\end{minted}

The script first invokes the \codeinline{create\_default\_params}
\href{https://docs.python.org/3/library/functions.html#classmethod}{classmethod}
which returns a \codeinline{Parameters} object, typically named
\codeinline{params} containing all default parameters.
Any modifications to simulation parameters is made after this step, to meet the
user's needs.  The simulation object is then instantiated by passing
\codeinline{params} as the only argument, typically named \codeinline{sim},
ready to start the iterations.

As demonstrated above, parameters are stored in an object of the class
\codeinline{Parameters}, which uses the
\href{https://fluiddyn.readthedocs.io/en/latest/generated/fluiddyn.util.paramcontainer.html}{\codeinline{ParamsContainer}}
API\footnote{See
\href{https://fluiddyn.readthedocs.io/en/latest/generated/fluiddyn.util.paramcontainer.html}{fluiddyn.readthedocs.io/en/latest/generated/fluiddyn.util.paramcontainer.html}}
of \fluidpack{dyn} package \cite{fluiddyn}.
Parameters for all possible modifications to initialization, preprocessing,
forcing, time-stepping, and output of the solvers is incorporated into the
object \codeinline{params} in a hierarchial manner.
Once initialized, the ``public'' (not hidden) API does not allow to add new
parameters to this object and only modifications are permitted.\footnote{Example
on modifying the parameters for a simple simulation:
\href{https://fluidsim.readthedocs.io/en/latest/examples/running-simul-onlineplot.html}{%
fluidsim.readthedocs.io/en/latest/examples/running\_simul.html}}

This approach is different from conventional solvers reliant on
text-based input files to specify parameters, which is less robust and can
cause the simulation to crash due to human errors during runtime.
A similar, but less readable approach to \codeinline{ParamsContainer} is
adopted by OpenFOAM which relies on \codeinline{dictionaries} to customize
parameters.  The \codeinline{params} object can be printed on the Python or
IPython console and explored interactively using tab-completion, and can be
loaded from and saved into XML and HDF5 files, thus being very versatile.

Note that the same simulation object is used for the plotting and post-processing
tasks.  During or after the execution of a simulation, a simulation object can be
created with the following code (to restart a simulation, one would rather use the
function \codeinline{fluidsim.load\_state\_phys\_file}.):
\begin{minted}[fontsize=\footnotesize]{python}
 from fluidsim import load_sim_for_plot

 # in the directory of the simulation
 sim = load_sim_for_plot()
 # or with the path of the simulation
 # sim = load_sim_for_plot("~/Sim_data/NS2D.strat_240x240_S8x8_2018-04-20_13-45-54")

 # to retrieve the value of a parameter
 print(f"viscosity = {sim.params.nu_2}")

 # to plot the space averaged quantities versus time
 sim.output.spatial_means.plot()
 # to load the corresponding data
 data = sim.output.spatial_means.load()

 # for a 2d plot of the variable "b"
 sim.output.phys_fields.plot("b", time=2)
 # to save a 2d animation
 sim.output.phys_fields.animate("b", tmin=1, tmax=5, save_file=True)
\end{minted}

\begin{figure}[htp]
\centering
\includegraphics[width=\linewidth]{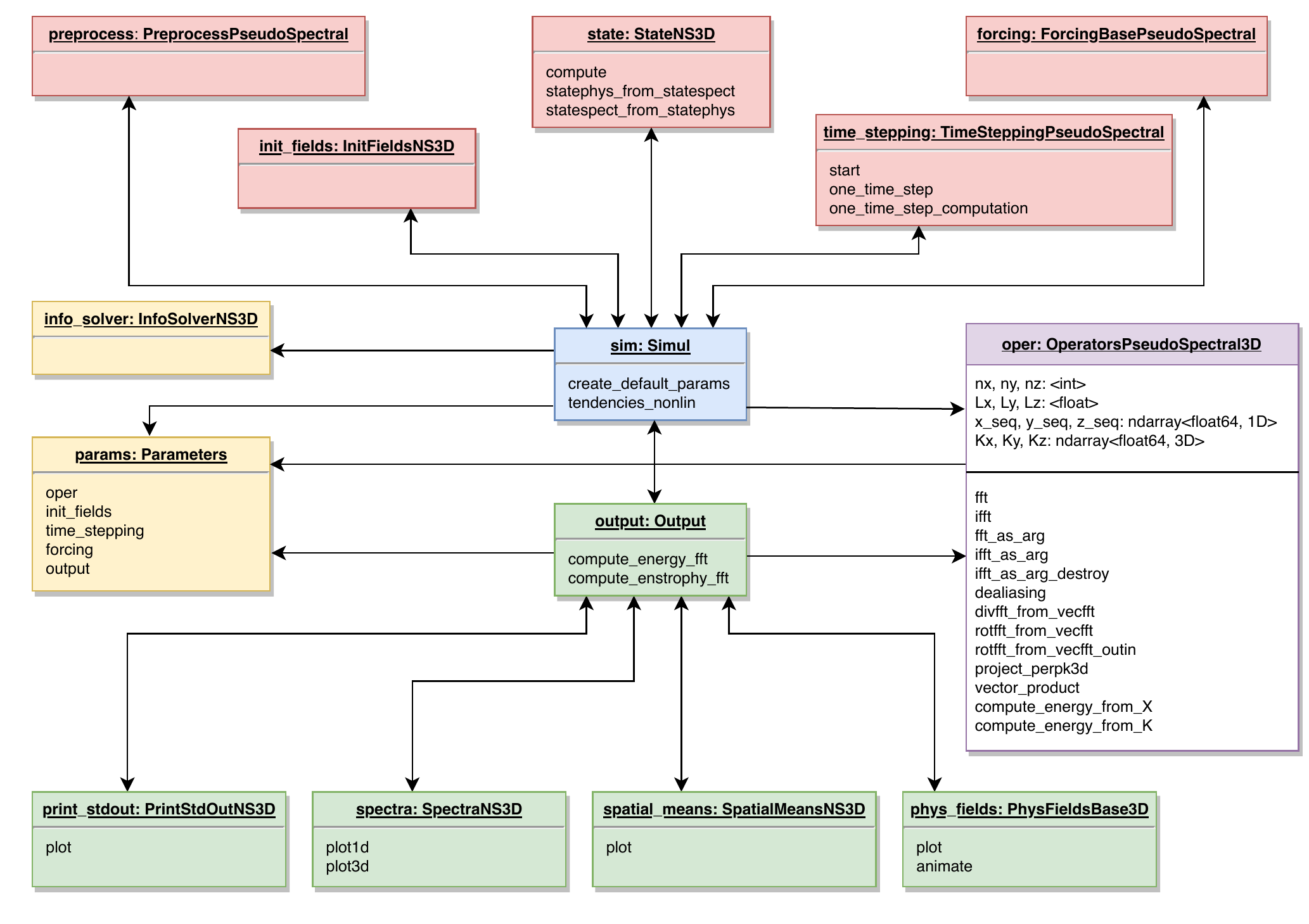}
\caption{%
\href{https://www.uml-diagrams.org}{UML diagram} of the simulation
object (\codeinline{sim}) for the solver
\codeinline{fluidsim.solvers.ns3d}.
Each block represents an object or instance of a class, and the object name
and the class name are written as headings. The solid arrows show how objects are
associated with each other.
Methods and variables of significance to the user are displayed in the body of
each object block.}
\label{fig:simul}
\end{figure}

Fig.~\ref{fig:simul} demonstrates how an user can access different objects and
its associated methods through the simulation object for the solver
\codeinline{fluidsim.solvers.ns3d}. Do note that the object and method names
are similar, if not same, for other solvers. The purpose of the objects are
listed below in the order of instantiation:

\begin{outline}
\1 \codeinline{sim.params}: A copy of the \codeinline{params} object supplied
by the user, which holds all information critical to run the simulation and
generating the output.
\1 \codeinline{sim.info\_solver}: Contains all the subclass information,
including the module the class belongs to.
\1 \codeinline{sim.info}: A union of the \codeinline{sim.info\_solver} and the
\codeinline{sim.params} objects.
\1 \codeinline{sim.oper}: Responsible for generating the grid, and for
pseudo-spectral numerical methods such as FFT, IFFT, dealiasing, divergence,
curl, random arrays, etc.
\1 \codeinline{sim.output}: Takes care of all on-the-fly post-processing
outputs and functions to load and plot saved output files. Different objects
are assigned with tasks of loading, plotting and sometimes computing:
\2 \codeinline{sim.output.print\_stdout}: the mean energies, time elapsed and
time-step of the simulation printed as console output.
\2 \codeinline{sim.output.phys\_fields}: the state variables in the physical
plane.  It relies on \codeinline{sim.state} to load or compute the variables
into arrays.
\2 \codeinline{sim.output.spatial\_means}: mean quantities such as energy,
enstrophy, forcing power, dissipation.
\2 \codeinline{sim.output.spectra}: energy spectra as line plots (i.e.\ as
functions of the module or a component of the wavenumber).
\2 \codeinline{sim.output.spect\_energy\_budg}: spectral energy budget by
calculating the transfer term.
\2 \codeinline{sim.output.increments}: structure functions from physical
velocity fields.
\1 \codeinline{sim.state}: Defines the names of the physical variables being
solved for and their spectral equivalents, along with all required variable
transformations.
Also includes high-level objects, aptly named \codeinline{sim.state.state\_phys}
and \codeinline{sim.state.state\_spect} to hold the arrays.
\1 \codeinline{sim.time\_stepping}: Generic numeric time-integration object
which dynamically determines the time-step using the CFL criterion for specific
solver and advances the state variables using Runge-Kutta method of order 2 or
4.
\1 \codeinline{sim.init\_fields}: Used only once to initialize all state variables
from a previously generated output file or with simple kinds of flow structures,
for example a dipole vortex, base flow with constant value for all gridpoints,
grid of vortices, narrow-band noise, etc.
\1 \codeinline{sim.forcing}: Initialized only when
\codeinline{params.forcing.enable} is set as \codeinline{True} and it computes
the forcing variables, which is added on to right-hand-side of the equations
being solved.
\1 \codeinline{sim.preprocess}: Adjusts solver parameters such as the magnitude
of initialized fields, viscosity value and forcing rates after all other
subclasses are initialized, and just before the time-integration starts.
\end{outline}

Such a modular organization of the solver's features has several advantages.
The most obvious one, will be the ease of maintaining the code base. As opposed
to a monolithic solver, modular codes are well separated and leads to less
conflicts while merging changes from other developers. Secondly, with this
implementation, it is possible to extend or replace a particular set of
features by inheriting or defining a new class.

Modular codes can be difficult to navigate and understand the connection between
objects and the classes in static languages. It is much less a problem with
Python where one can easily decipher this information from object attributes or
using IPython's dynamic object information feature\footnote{See
\href{https://ipython.readthedocs.io/en/stable/interactive/%
reference.html\#dynamic-object-information}{ipython.readthedocs.io}.}.
Now, \fluidpack{sim} goes one step further and one can effortlessly print the
\codeinline{sim.info\_solver} object in the Python / IPython console, to get this
information.  A truncated example of the output is shown below.

\begin{minted}[fontsize=\footnotesize]{python}
>>> sim.info_solver
<fluidsim.solvers.ns3d.solver.InfoSolverNS3D object at 0x7fb6278263c8>

<solver class_name="Simul" module_name="fluidsim.solvers.ns3d.solver"
        short_name="ns3d">
  <classes>
    <Operators class_name="OperatorsPseudoSpectral3D"
               module_name="fluidsim.operators.operators3d"/>

    <State class_name="StateNS3D" keys_computable="['rotz']"
           keys_linear_eigenmodes="['rot_fft']" keys_phys_needed="['vx', 'vy',
           'vz']" keys_state_phys="['vx', 'vy', 'vz']"
           keys_state_spect="['vx_fft', 'vy_fft', 'vz_fft']"
           module_name="fluidsim.solvers.ns3d.state"/>

    <TimeStepping class_name="TimeSteppingPseudoSpectralNS3D"
                  module_name="fluidsim.solvers.ns3d.time_stepping"/>

    <InitFields class_name="InitFieldsNS3D"
                module_name="fluidsim.solvers.ns3d.init_fields">
      <classes>
        <from_file class_name="InitFieldsFromFile"
                   module_name="fluidsim.base.init_fields"/>

        <from_simul class_name="InitFieldsFromSimul"
                    module_name="fluidsim.base.init_fields"/>

        <in_script class_name="InitFieldsInScript"
                   module_name="fluidsim.base.init_fields"/>

        <constant class_name="InitFieldsConstant"
                  module_name="fluidsim.base.init_fields"/>

	<!--truncated output-->

  </classes>

</solver>
\end{minted}

Note that while the 3D Navier-Stokes solver relies on some generic \emph{base
classes}, such as \codeinline{OperatorsPseudoSpectral3D} and
\codeinline{TimeSteppingPseudoSpectral}, shared with other solvers; for other
purposes there are \emph{solver specific classes}.  The latter is often inherited
from the base classes in \codeinline{fluidsim.base} or classes available in other
solvers --- this made possible by the use of an object-oriented approach.  This is
particularly advantageous while extending existing features or creating new
solvers to use
\href{https://docs.python.org/3/tutorial/classes.html#inheritance}{%
class inheritance}.

\subsection*{Performance}

%
%

Performance of a code can be carefully measured by three different program
analysis methods: profiling, micro-benchmarking and scalability analysis.
\href{https://en.wikipedia.org/wiki/Profiling_(computer_programming)}{Profiling}
traces various function calls and records the cumulative time consumed and the
number of calls for each function.  Through profiling, we shed light on what
part of the code consumes the lion's share of the computational time and
observe the impact as number of processes and MPI communications increase.
\href{https://en.wiktionary.org/wiki/microbenchmark}{Micro-benchmarking} is the
process of timing a specific portion of the code and comparing different
implementations.  The aspect is addressed in greater detail in the companion
paper on \fluidpack{fft} \citep{fluidfft}.
On the other hand,
\href{https://en.wikipedia.org/wiki/Scalability#Performance_tuning_versus_hardware_scalability}{a
scalability study} measures how the speed of the code improves when it is deployed
with multiple CPUs operating in parallel.  To do this, the walltime required by
the whole code is measured to complete a certain number of iterations, given a
problem size.
Finally, performance can also be studied by comparing different codes on
representative problems. Since such comparisons should not focus only on
performance, we present a comparison study in a separate section.

\begin{table}[h]
\centering
\begin{tabular}{l p{8cm}}
  \toprule
  Cluster & Beskow (Cray XC40 system with Aries interconnect) \\
  CPU &  Intel Xeon CPU E5--2695v4, 2.1GHz \\
  Operating System & SUSE Linux Enterprise Server 11, Linux Kernel 3.0.101\\
  No.\ of cores per nodes used & 32 \\
  Maximum no.\ of nodes used & 32 (2D cases), 256 (3D cases) \\
  Compilers & CPython 3.6.5, Intel C++ Compiler (\pack{icpc}) 18.0.0 \\
  Python packages & \fluidpack{dyn} 0.2.3, \fluidpack{fft} 0.2.3,
  \fluidpack{sim} 0.2.1, \pack{numpy} (OpenBLAS) 1.14.2, \pack{Cython} 0.28.1,
  \pack{mpi4py} 3.0.0, \pack{pythran} 0.8.5 \\

  \bottomrule
\end{tabular}
\caption{Specifications of the supercomputing cluster and software used for profiling and benchmarking.}
\label{tab:specs}

\end{table}

The profiling and scaling tests were conducted in a supercomputing cluster of
the Swedish National Infrastructure for Computing (SNIC) namely Beskow (PDC,
Stockholm). Relevant details regarding software and hardware which affect
performance are summarised in Table~\ref{tab:specs}. Note that no
hyperthreading was used while carrying out the studies.
The code comparisons were made on a smaller machine.
Results from the three analyses are systematically studied in the following
sections.

\subsubsection*{Profiling}

\begin{figure}[htp]
\centering
\includegraphics[width=\linewidth]{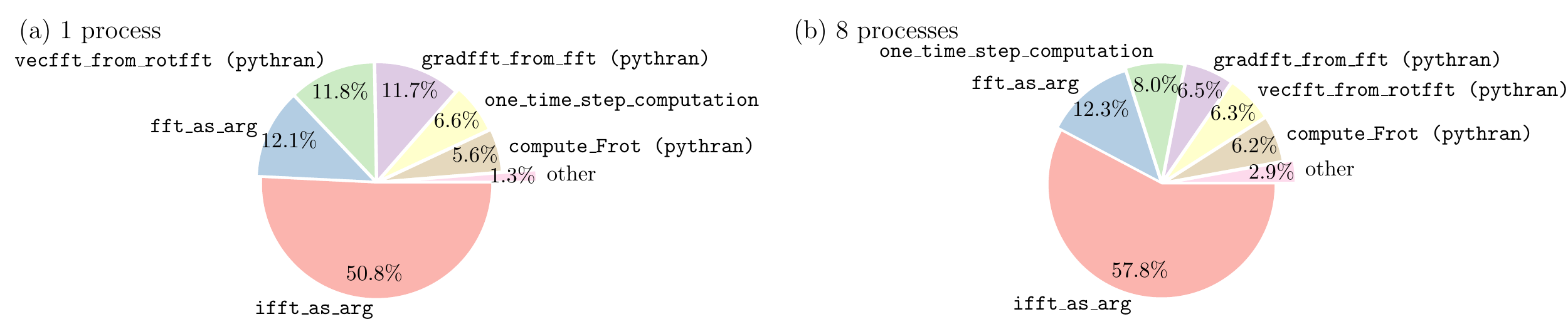}
\caption{Profiling analysis of the 2D Navier-Stokes
(\codeinline{fluidsim.solvers.ns2d}) solver using a grid sized $1024\times1024$
(a) in sequential with \codeinline{fft2d.with\_fftw1d} operator and (b) with 8
processes with \codeinline{fft2d.mpi\_with\_fftwmpi2d}
operator.}\label{fig:profile2d}
\end{figure}

\begin{figure}[htp]
\centering
\includegraphics[width=\linewidth]{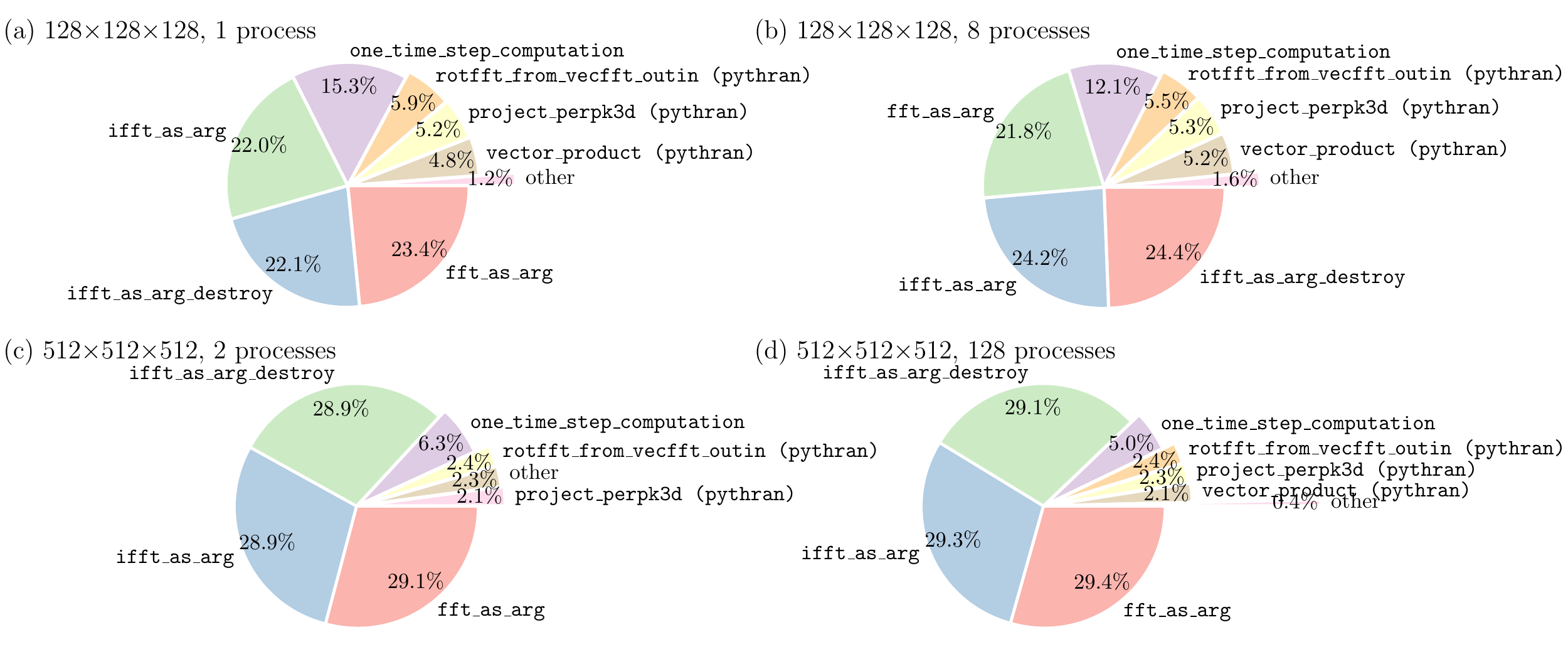}
\caption{Profiling analysis of the 3D Navier-Stokes
(\codeinline{fluidsim.solvers.ns3d}) solver.
Top row: grid sized $128\times128\times128$ solved (a) sequentially using
\codeinline{fft3d.with\_fftw3d} operator and (b) with 8 processes using
\codeinline{fft3d.mpi\_with\_fftwmpi3d} operator.
Bottom row: grid sized $512\times512\times512$ using
\codeinline{fft3d.mpi\_with\_fftwmpi3d} operator (c) with 2 processes and
(d) with 128 processes.}
\label{fig:profile3d}
\end{figure}

It is straightforward to perform profiling with the help of the
\href{https://docs.python.org/3/library/profile.html}{cProfile} module, available
in the Python standard library.
For \fluidpack{sim}, this module has been conveniently packaged into a command
line utility, \codeinline{fluidsim-profile}.
Here, we have analyzed both 2D and 3D Navier-Stokes solvers in Beskow, and
plotted the results in Fig.~\ref{fig:profile2d} and Fig.~\ref{fig:profile3d}
respectively. Functions which consume less than $2\%$ of the total time are
displayed within a single category, \emph{other}.


In Fig.~\ref{fig:profile2d} both sequential and parallel profiles of the 2D
Navier-Stokes solver shows that majority of time is spent in inverse and forward
FFT calls (\codeinline{ifft\_as\_arg} and \codeinline{fft\_as\_arg}). For the
sequential case, approximately 0.14\% of the time is spent in pure Python
functions, i.e.\ functions not built using \pack{Cython} and \pack{Pythran}.
\pack{Cython} extensions are responsible for interfacing with FFT operators and
also for the time-step algorithm.  \pack{Pythran} extensions are used to translate
most of the linear algebra operations into optimized, statically compiled
extensions.
We also see that only 1.3\% of the time is not spent in the main six functions
(category \emph{other} in the figure).
With 8 processes deployed in parallel, time spent in pure Python function
increases to 1.1\% of the total time.
These results show that during the optimization process, we have to focus on a
very small number of functions.

From Fig.~\ref{fig:profile3d} it can be shown that, for the 3D Navier-Stokes
solver for all cases majority of time is attributed to FFT calls. The overall time
spent in pure Python function range from 0.001\% for $512^3$ grid points and 2
processes to 0.46\% for $128^3$ grid points and 8 processes.
This percentage tends to increase with the number of processes used since the real
calculation done in compiled extensions take less time.
This percentage is also higher for the coarser resolution for the same reason.
However, the time in pure Python remains for all cases largely negligible compared
to the time spent in compiled extensions.

\subsubsection*{Scalability}

Scalability can be quantified by speedup $S$ which is a measure of the time taken
to complete $N$ iterations for different number of processes, $n_p$.  We shall
refrain from comparing sequential runs in this context, since the operators used
for the sequential mode differ from the parallel mode, especially the FFT
class. Speedup is formally defined here as:

\begin{equation}
S_\alpha(n_p) = \frac
{[\mathrm{Time\ elapsed\ for\ } N \mathrm{\ iterations\ with\ }n_{p,\min}\mathrm{\ processes}]_{\mathrm{fastest}}
\times n_{p,\min}}
{[\mathrm{Time\ elapsed\ for\ } N \mathrm{\ iterations\ with\ } n_p \mathrm{\
processes}]_\alpha}
\label{eq:speedup}
\end{equation}
where $n_{p,\min}$ is the minimum number of processes employed for a specific
array size and hardware, $\alpha$ denotes the FFT class used and ``fastest''
corresponds to the fastest result among various FFT classes.
In addition to number of processes, there is another important parameter, which
is the size of the problem; in other words, the number of grid points used to
discretize the problem at hand.
In \emph{strong scaling} analysis, we keep the global grid-size fixed and
increase the number of processes.

Ideally, this should yield a speedup which increases linearly with number of
processes.  Realistically, as the number of processes increase, so does the
number of MPI communications, contributing to some latency in the overall time
spent and thus resulting in less than ideal performance.
Also, as shown by profiling in the previous section, majority of the time is
consumed in making forward- and inverse-FFT calls, an inherent bottleneck of
the pseudo-spectral approach. The FFT function calls are the source of most of
the MPI calls during runtime, limiting the parallelism.

\subsubsection*{2D benchmarks}\label{sec:bench2d}

The Navier-Stokes 2D solver (\codeinline{fluidsim.solvers.ns2d}) solving an
initial value problem (with random fields) was chosen as the test case for strong
scaling analysis here. The physical grid was discretized
with $1024\times1024$ and $2048\times2048$ points.
Fourth-order Runge-Kutta (RK4) method with a constant time-step was used for
time-integration.
File input-output and the forcing term has been disabled so as to measure the
performance accurately.  The test case is then executed for 20 iterations.
The time elapsed was measured just before and after the
\codeinline{sim.time\_stepping.start()} function call, which was then utilized
to calculate the average walltime per iteration and speedup.
This process is repeated for two different FFT classes provided by
\fluidpack{fft}, viz. \codeinline{fft2d.mpi\_with\_fftw1d} and
\codeinline{fft2d.mpi\_with\_fftwmpi2d}.

\begin{figure}[htp]
\centering
\includegraphics[width=\linewidth]{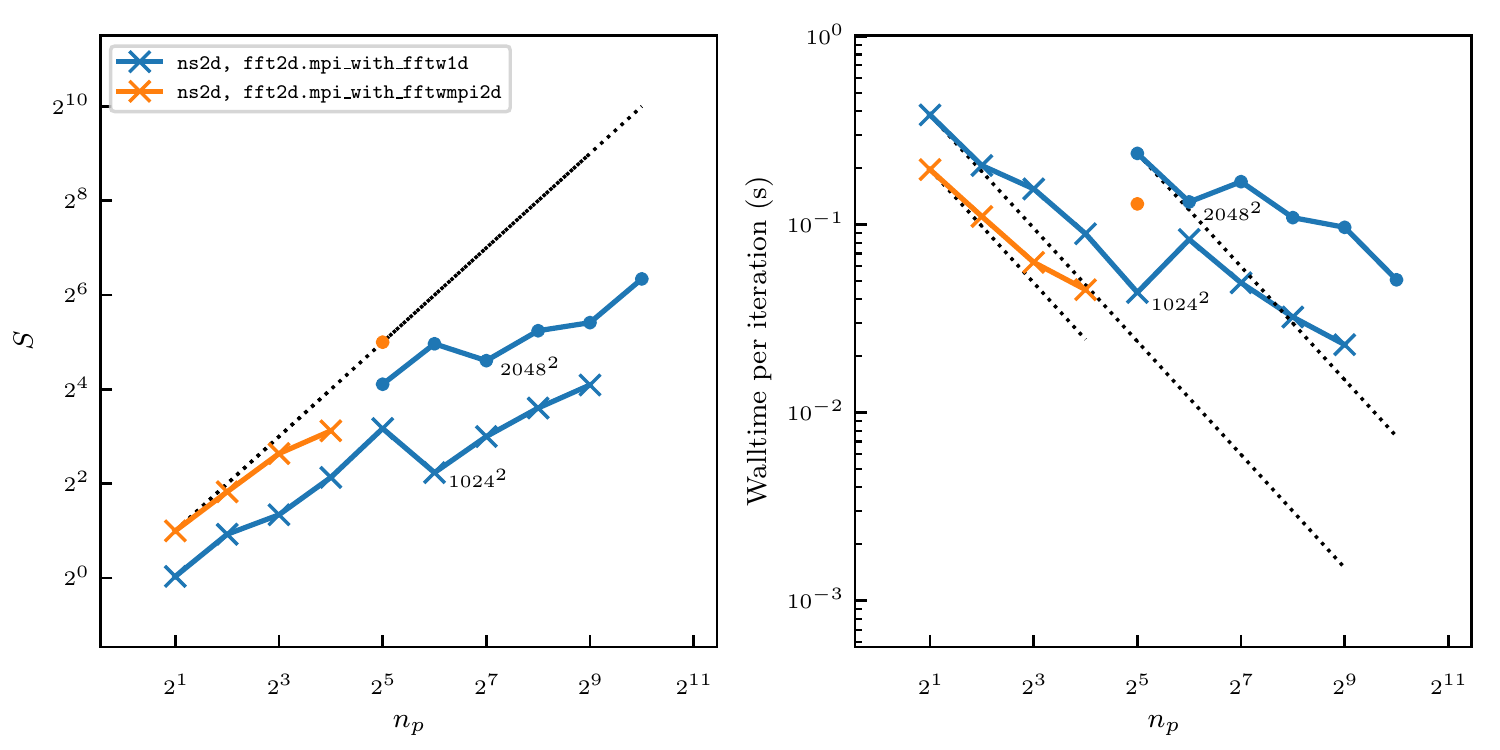}
\caption{Strong scaling benchmarks of the 2D Navier-Stokes
(\codeinline{fluidsim.solvers.ns2d}) solver. The number of cores $n_p$ goes from 2
to $2^{10} = 1024$. Crosses and dots correspond to $1024\times1024$ and
$2048\times2048$ grid points, respectively.}
\label{fig:strong2d}
\end{figure}

In Fig.~\ref{fig:strong2d} we have analyzed the strong scaling speedup $S$ and
walltime per iteration. The fastest result for a particular case is assigned
the value $S=n_p$ as mentioned earlier in Eq.~\ref{eq:speedup}. Ideal speedup
is indicated with a dotted black line and it varies linearly with number of
processes.  We notice that for the $1024\times1024$ case there is an assured
increasing trend in speedup for intra-nodes computation.
Nevertheless, when this test case is solved with over a node ($n_p > 32$); the
speedup drops abruptly. While it may be argued that the speedup is impacted by
the cost of inter-node MPI communications via network interfaces, that is not
the case here. This is shown by speedup for the $2048\times2048$ case, where
speedup increases from $n_p = 32$ to $64$, after which it drops again. It is thus
important to remember that a decisive factor in pseudo-spectral simulations is
the choice of the grid size, both global and local (per-process), and for certain
shapes the FFT calls can be exceptionally fast or vice-versa.

%

From the above results, it may also be inferred that superior performance is
achieved through the use of \codeinline{fft2d.mpi\_with\_fftwmpi2d} as the FFT method. The
\codeinline{fft2d.mpi\_with\_fftw1d} method serves as a fallback option when
either FFTW library is not compiled using MPI bindings or the domain
decomposition results in zero-shaped arrays, which is a known issue with the current
version of \fluidpack{sim} and requires further development.

To the right of Fig.~\ref{fig:strong2d}, the real-time or walltime required to
perform a single iteration in seconds is found to vary inversely proportional
to the number of processes, $n_p$. The walltime per iteration ranges from
$0.195$ to $0.023$ seconds for the $1024\times1024$ case, and from
$0.128$ to $0.051$ seconds for the $2048\times2048$ case. Thus it is indeed
feasible and scalable to use this particular solver.

\subsubsection*{3D benchmarks}

\begin{figure}[htp]
\centering
\includegraphics[width=\linewidth]{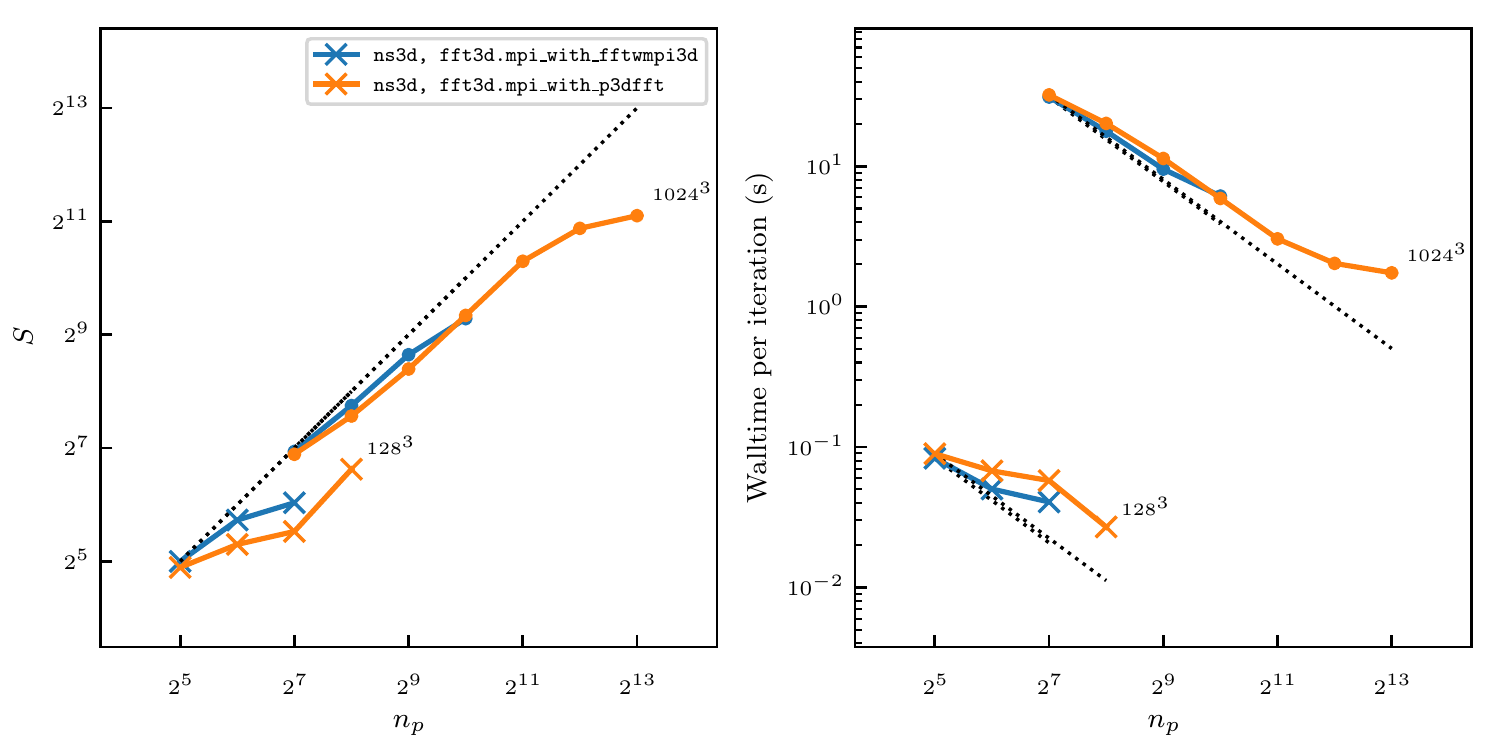}
\caption{Strong scaling benchmarks of the 3D Navier-Stokes
(\codeinline{fluidsim.solvers.ns3d}) solver in Beskow. The number of cores $n_p$
goes from $2^5 = 32$ to $2^{13} = 8192$. Crosses and dots correspond to $128^3$
and $1024^3$ grid points, respectively.}
\label{fig:strong3d_beskow}
\end{figure}

Using a similar process as described in the previous section,
the Navier-Stokes 3D solver (\codeinline{fluidsim.solvers.ns3d}) is chosen to
perform 3D benchmarks.
As demonstrated in Fig.~\ref{fig:strong3d_beskow} two physical global grids
with $128\times128\times128$ and $1024\times1024\times1024$ are used to
discretize the domain.
Other parameters are identical to what was described for the 2D benchmarks.

Through \fluidpack{fft}, this solver has four FFT methods at disposal:

\begin{itemize}
 \item \codeinline{fft3d.mpi\_with\_fftw1d}
 \item \codeinline{fft3d.mpi\_with\_fftwmpi3d}
 \item \codeinline{fft3d.mpi\_with\_p3dfft}
 \item \codeinline{fft3d.mpi\_with\_pfft}
\end{itemize}

The first two methods implements a 1D or \emph{slab} decomposition, i.e.\ the
processes are distributed over one index of a 3D array. And the last two
methods implement a 2D or \emph{pencil} decomposition. For the sake of clarity,
we have restricted this analysis to the fastest FFT method of the two types in
this configuration, viz. \codeinline{fft3d.mpi\_with\_fftwmpi3d} and
\codeinline{fft3d.mpi\_with\_p3dfft}. A more comprehensive study of the
performance of these FFT methods can be found in \citet{fluidfft}.

In Fig.~\ref{fig:strong3d_beskow} the strong scaling speedup and walltime per
iteration are plotted from 3D benchmarks in Beskow.
The analysis here is limited to single-node and inter-node performance.
For both grid-sizes analyzed here, the \codeinline{fft3d.mpi\_with\_fftwmpi3d}
method is the fastest of all methods but limited in scalability because of the
1D domain decomposition strategy.  To utilize a large number of processors, one
requires the 2D decomposition approach. Also, note that for the
$1024\times1024\times1024$ case, a single-node measurement was not possible as
the size of the arrays required to run the solvers exceeds the available
memory. For the same case, a speedup reasonably close to linear variation is
observed with \codeinline{fft3d.mpi\_with\_p3dfft}.
It is also shown that the walltime per iteration improved from
$0.083$ to $0.027$ seconds for the $128\times128\times128$ case, and from
$31.078$ to $2.175$ seconds for the $1024\times1024\times1024$ case.

\subsection*{CFD pseudo-spectral code comparisons}



As a general CFD framework, \fluidpack{sim} could be compared to OpenFOAM (a CFD
framework based on finite-volume methods).
However, in contrast to OpenFOAM, the current version of \fluidpack{sim} is highly
specialized in pseudo-spectral Fourier methods and it is not adapted for
industrial CFD.

In this subsection, we compare \fluidpack{sim} with three other open-source CFD
pseudo-spectral codes\footnote{For the sake of conciseness, we limit this
comparison to only four codes. We have also found the Julia code
\href{https://github.com/FourierFlows/FourierFlows.jl}{FourierFlows.jl} to
demonstrate interesting performance for 2D sequential runs, but without support
for 3D cases and MPI parallelization.}:
\begin{itemize}
\item \href{http://dedalus-project.org/}{Dedalus} \citep{burns_dedalus} is ``a
flexible framework for spectrally solving differential equations''. It is very
versatile and the user describes the problem to be solved symbolically.
This approach is very different than the one of \fluidpack{sim}, where the
equations are described with simple \Numpy code.  There is no equivalent of the
\fluidpack{sim} concept of a ``solver'', i.e.\ a class corresponding to a set of
equations with specialized outputs (with the corresponding plotting methods).  To
run a simulation with Dedalus, one has to describe the problem using mathematical
equations.  This can be very convenient because it is very versatile and it is not
necessary to understand how Dedalus works to define a new problem. However, this
approach has also drawbacks:
\begin{itemize}
\item Even for very standard problems, one needs to describe the problem in the
launching script.
\item There is a potentially long initialization phase during which Dedalus
processes the user input and prepares the ``solver''.
\item Even when a user knows how to define a problem symbolically, it is not
simple to understand how the problem is solved by Dedalus and how to interact with
the program with Python.
\item Since solvers are not implemented out-of-the-box in Dedalus,
specialized forcing scheme or outputs are absent. For example, the user has to
implement the computation, saving and plotting of standard outputs like energy
spectra.
\end{itemize}

\item \href{https://github.com/spectralDNS/spectralDNS}{SpectralDNS}
\citep{mortensen_spectraldns2016} is a ``high-performance pseudo-spectral
Navier-Stokes DNS solver for triply periodic domains. The most notable feature of
this solver is that it is written entirely in Python using \Numpy, MPI
for Python (\pack{mpi4py}) and \pack{pyFFTW}.''

Therefore, SpectralDNS is technically very similar to \fluidpack{sim}.
Some differences are that SpectralDNS has no object oriented API, and that the user
has to define output and forcing in the launching script\footnote{See
\href{https://github.com/spectralDNS/spectralDNS/tree/master/demo}{the demo
scripts of SpectralDNS}.}, which are thus usually much longer than for
\fluidpack{sim}.
Moreover, the parallel Fourier transforms are done using the Python package
\href{https://github.com/spectralDNS/mpiFFT4py}{\pack{mpiFFT4py}}, which is only
able to use the FFTW library and not other libraries as with \fluidpack{fft}
\citep[][]{fluidfft}.

\item \href{https://bitbucket.org/paugier/ns3d}{NS3D} is a highly efficient
pseudo-spectral Fortran code.
It has been written in the laboratory
\href{https://www.ladhyx.polytechnique.fr}{LadHyX} and used for several studies
involving simulations (in 3D and in 2D) of the Navier-Stokes equations under the
Boussinesq approximation with stratification and system rotation
\cite[][]{DeloncleBillantChomaz2008}.  It is in particular specialized in
stability studies \cite[][]{BillantDeloncleChomazOtheguy2010}.
NS3D has been highly optimized and it is very efficient for sequential and
parallel simulations (using MPI and OpenMP).  However, the parallelization is
limited to 1D decomposition for the FFT \cite[][]{fluidfft}.
Another weakness compared to \fluidpack{sim} is that NS3D uses simple binary files
instead of HDF5 and NetCDF4 files for \fluidpack{sim}.  Therefore, visualization
programs like Paraview or Visit cannot load NS3D data.

As with many Fortran codes, Bash and Matlab are used for launching and
post-processing, respectively.
In terms of user experience, this can be a drawback compared to the coherent
framework \fluidpack{sim} for which the user works only with Python.

In contrast to the framework \fluidpack{sim} for which it is easy to define a new
solver for a new set of equations, NS3D is specialized in solving the
Navier-Stokes equations under the Boussinesq approximation. Using NS3D to solve a
new set of equations would require very deep changes in many places in the code.

\end{itemize}

For quantitative comparisons and for the sake of simplicity, we limit
ourselves to compare only sequential runs.  We have already discussed in
detail, the issue of the scalability of pseudo-spectral codes based on Fourier
transforms in the previous section and in the companion paper \citep{fluidfft}.
We compare the code with a very simple and standard task, running a solver for
ten time steps with the Runge-Kutta 4 method.
Note that Dedalus does not implement the standard fully explicit RK4
method\footnote{See
\href{https://bitbucket.org/dedalus-project/dedalus/issues/38/%
slow-simulation-ns2d-over-a-biperiodic}{the Dedalus issue 38.}}. Thus for
Dedalus, we use the most similar time stepping scheme available, RK443, a
4-stage, third-order mixed implicit-explicit scheme described in
\citet{ascher1997implicit}.
Note that in the other codes, part of the linear terms are also treated
implicitly.
Also note that in several cases, the upper bound of time step is not first
limited by the stability of the time scheme, rather by other needs (to
resolve the fastest wave, accuracy, etc.), so these benchmarks are
representative of elapsed time for accurate real-life simulations.

\paragraph{Bi-dimensional simulations.}

\begin{table}
\centering
\begin{tabular}{lrrrr}
\hline
          &   \fluidpack{sim} &   Dedalus &   SpectralDNS &   NS3D \\
\hline
 512$^2$  &              0.54 &      8.01 &          0.92 &   0.82 \\
 1024$^2$ &              2.69 &     43.00 &          3.48 &   3.96 \\
\hline
\end{tabular}
\caption{Elapsed times (in seconds) for ten RK4 time steps for two bidimensional
cases and the four CFD codes.}
\label{table:compare}
\end{table}

We first compare the elapsed times for two resolutions (512$^2$ and 1024$^2$) over
a bi-dimensional space.  The results are summarized in Table~\ref{table:compare}.
The results are consistent for the two resolutions.  \fluidpack{sim} is the
fastest code for these cases.  Dedalus is more than one order of magnitude slower
but as discussed earlier, the time stepping method is different. Also note
that Dedalus has more been optimized for bounded domains with Chebyshev
methods.
The two other codes SpectralDNS and NS3D have similar performance: slightly slower
than \fluidpack{sim} and much faster than Dedalus.
Surprisingly, the Fortran code NS3D is slower (47\%) than the Python code
\fluidpack{sim}.  This can be explained by the fact that there is no specialized
numerical scheme for the 2D case in NS3D, so that more FFTs have to be performed
compared to SpectralDNS and \fluidpack{sim}.
This highlights the importance of implementing a well-adapted algorithm for
a class of problems, which is much easier with a highly modular code as
\fluidpack{sim} than with a specialized code as NS3D.

\paragraph{Tri-dimensional simulations.}

We now compare the elapsed times for ten RK4 time steps for a tri-dimensional case
with a resolution 128$^3$.
Dedalus is slow and does not seem to be adapted for this case so we do not give
exact elapsed time for this code.
SpectralDNS is slightly slower (11.55~s) than the two other codes (9.45~s for
\fluidpack{sim} and 9.52~s for NS3D). This difference is mainly explained by the
slower FFTs for SpectralDNS.

\begin{figure}[htp]
\centering
\includegraphics[width=\linewidth]{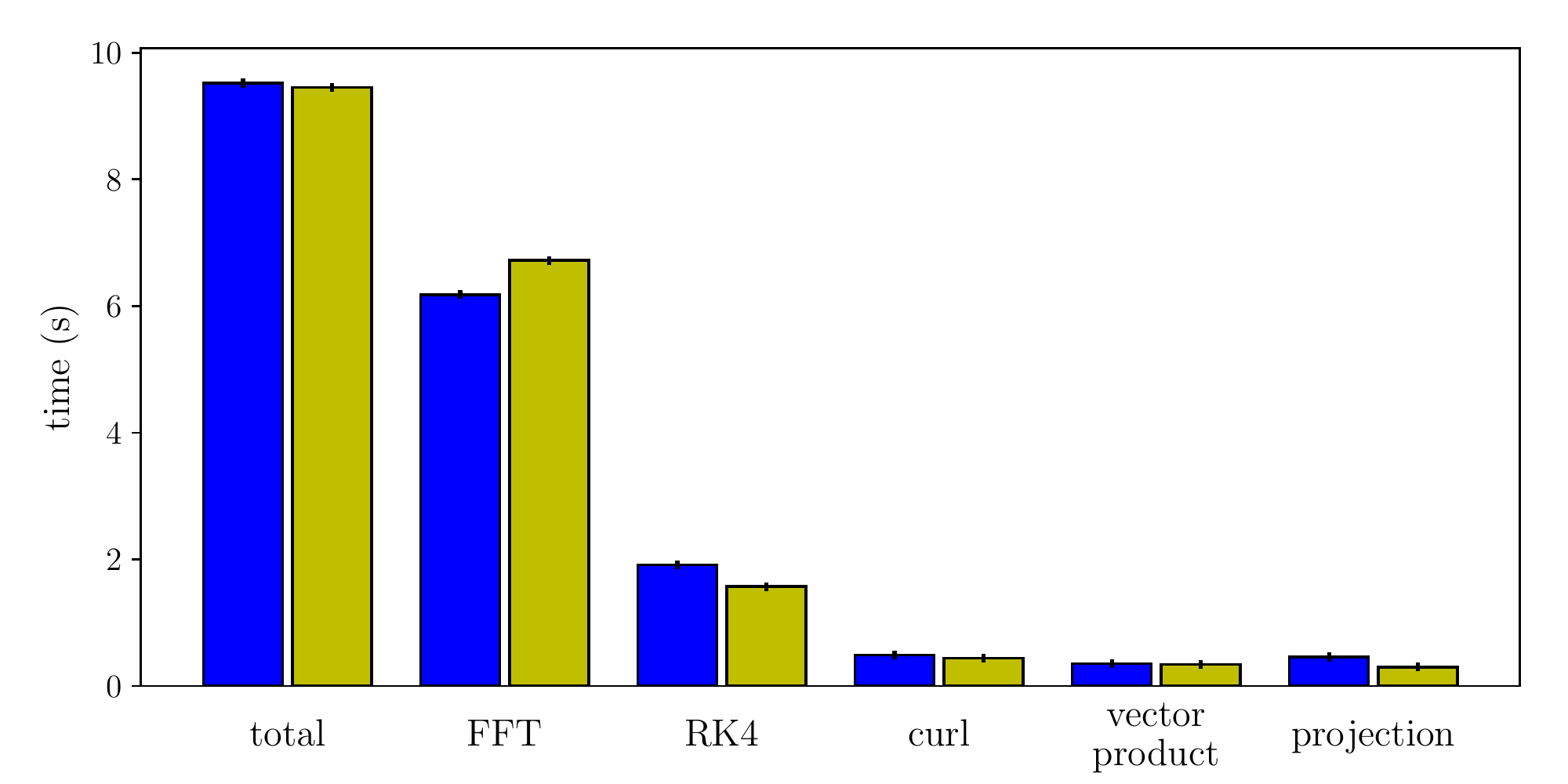}
\caption{Comparison of the execution times for a 3D case (128$^3$, 10 time steps)
between NS3D (blue bars) and \codeinline{fluidsim.solvers.ns3d} (yellow bars).
The first two bars correspond to the total time and the others to the main tasks
in terms of time consumption, namely FFT, Runge-Kutta 4, curl, vector product and
``projection''. }
\label{fig:compare:with:ns3d}
\end{figure}

Fig.~\ref{fig:compare:with:ns3d} presents a more detailed comparison between NS3D
(blue bars) and \fluidpack{sim} (yellow bars).
The total elapsed times is mainly spent in five tasks: FFTs, Runge-Kutta 4, curl,
vector product and ``projection''. The times spent to perform these tasks are
compared for the two codes.

We see that FFTs in NS3D are very fast: the FFT execution is 0.55 s longer for
\fluidpack{sim} (nearly 9\% longer). This difference is especially significant for
sequential run for which there is no communication cost involved in the FFT
computation, thus making it the least favorable case for \fluidpack{sim}.
Indeed, MPI communications are input-output bounded tasks which are not faster
in Fortran than in Python.

This difference can partially be explained by the fact that in NS3D, all FFTs are
inplace (so the input can be erased during the transform). On one hand, this
choice is good for performance and for a lower memory consumption.  On the other
hand, since the same variables are used to store the fields in real and in Fourier
spaces, it makes the code harder to write, to understand and to modify.  Since
memory consumption in clusters is much less of a problem than in the
past and that code simplicity is highly important for a framework like
\fluidpack{sim}, we choose to use out-of-place FFTs in \fluidpack{sim}.
Another factor is that the flag FFTW\_PATIENT is used in NS3D which leads to very
long initialization and sometimes faster FFTs. Since we did not see significant
speed-up by using this flag in \fluidpack{sim} and that we also care about
initialization time, this flag is not used and we prefer to use the flag
FFTW\_MEASURE, which usually leads to similar performance.

Time stepping in NS3D is significantly slower than in \fluidpack{sim} (0.34 s
$\simeq$ 20 \% slower). We did not find a performance issue in NS3D.
The linear operators are slightly faster in \fluidpack{sim} than in the Fortran
code NS3D.  This is because this corresponds to Pythran functions written with
explicit loops \cite[see][]{fluidfft}.

Although the FFTs are a little bit faster for NS3D, the total time is slightly
smaller (less than 1\% of the total time) for \fluidpack{sim} for this case.

These examples do not prove that \fluidpack{sim} is always faster than NS3D or
is as fast as any very well optimized Fortran codes. However, they do
demonstrate that our very high-level and modular Python code is very efficient
and is not slower than a well-optimized Fortran code.

\section*{Quality control}


\fluidpack{sim} also packages unittests to go alongside the related
modules. Throughout the development process it is made sure that all tests pass
on priority to ensure that new changes to package does not damage existing
functionality.

It is also important to quantify the efficacy of the tests, and this
is done by calculating the code coverage. Code coverage is the ratio of the
number of lines tested by unittests over the total number of lines in the whole
package. For the present version of \fluidpack{sim} the code coverage is
valued at approximately 60\%.
For \fluidpack{sim}, the code coverage results are displayed at
\href{https://codecov.io/bb/fluiddyn/fluidsim}{Codecov}.

We also try to follow a consistent code style as recomended by PEP (Python
enhancement proposals) 8 and 257. This is also inspected using lint
checkers such as \codeinline{flake8} and \codeinline{pylint} among the
developers. The code is regularity cleaned up using the Python code formatter
\codeinline{black}.

All the above quality control techniques are implemented within the continuous
testing solutions, Travis CI and Bitbucket Pipelines. Instructions on how to
run unittests, coverage and lint tests are included in the documentation.

\section*{(2) Availability}
\vspace{0.5cm}
\section*{Operating system}

Any POSIX based OS, such as GNU/Linux and macOS.

\section*{Programming language}

Python 2.7, 3.4 or above.

\section*{Dependencies}

\begin{itemize}
\item {\bf Minimum:} \fluidpack{dyn}, \Numpy, \pack{h5netcdf}, \fluidpack{fft}
\cite[and FFT libraries, see][]{fluidfft}.
\item {\bf Optional:} \Scipy, \pack{mpi4py}, \pack{Cython} and
\pack{Pythran}, \pack{pulp}.
\end{itemize}

\section*{List of contributors}

\begin{itemize}
\item Ashwin Vishnu Mohanan (KTH): Development of the shallow water equations
solver, \codeinline{fluidsim.solvers.sw1l}, testing, continuous integration,
code coverage and documentation.
\item Cyrille Bonamy (LEGI): Extending the sub-package
\codeinline{fluidsim.operators.fft} (currently deprecated) into a
dedicated package, \fluidpack{fft} used by \fluidpack{sim} solvers.
\item Miguel Calpe (LEGI): Development of the 2D Boussinesq equation solver,
\codeinline{fluidsim.solvers.ns2d.strat}.
\item Pierre Augier (LEGI): Creator of \fluidpack{sim} and FluidDyn project,
developer of majority of the modules and solvers, future-proofing with Python 3
compatibility and documentation.
\end{itemize}

\section*{Software location:}

{\bf Archive}

\begin{description}[noitemsep,topsep=0pt]
\item[Name:] PyPI
\item[Persistent identifier:] https://pypi.org/project/fluidsim
\item[Licence:] CeCILL, a free software license adapted to both international
and French legal matters, in the spirit of and retaining compatibility with the
GNU General Public License (GPL).
\item[Publisher:] Pierre Augier
\item[Version published:] 0.2.2
\item[Date published:] 02/07/2018
\end{description}

{\bf Code repository}

\begin{description}[noitemsep,topsep=0pt]
\item[Name:] Bitbucket
\item[Persistent identifier:] https://bitbucket.org/fluiddyn/fluidsim
\item[Licence:] CeCILL
\item[Date published:] 2015
\end{description}

{\bf Emulation environment}

\begin{description}[noitemsep,topsep=0pt]
\item[Name:] Docker
\item[Persistent identifier:] https://hub.docker.com/r/fluiddyn/python3-stable
\item[Licence:] CeCILL-B, a BSD compatible French licence.
\item[Date published:] 02/10/2017
\end{description}

\section*{Language}

English

\section*{(3) Reuse potential}


\fluidpack{sim} can be used in research and teaching to run numerical simulations
with existing solvers.
Its simplicity of use and its plotting capacities make it particularly adapted for
teaching.
\fluidpack{sim} is used at LEGI and at KTH for studies on geophysical turbulence
\cite[see for example][]{LindborgMohanan2017}.
Since it is easy to modify any characteristics of the existing solvers or to build
new solvers, \fluidpack{sim} is a good tool to carry out other types of
simulations for academic studies.
The qualities and advantages of \fluidpack{sim} (integration with the Python
ecosystem and the FluidDyn project, documentation, reliability --- thanks to unittests
and continuous integration ---, versatility, efficiency and scalability) make us think
that \fluidpack{sim} can become a true collaborative code.

There is no formal support mechanism. However, bug reports can be submitted at
the \href{https://bitbucket.org/fluiddyn/fluidsim/issues}{Issues page on
Bitbucket}. Discussions and questions can be aired on instant messaging
channels in Riot (or equivalent with Matrix protocol)\footnote{
\url{%
  https://matrix.to/\#/\#fluiddyn-users:matrix.org}}
or via IRC protocol on Freenode at \codeinline{\#fluiddyn-users}. Discussions
can also be exchanged via the official mailing list\footnote{
\url{https://www.freelists.org/list/fluiddyn}}.

\section*{Acknowledgements}


Ashwin Vishnu Mohanan could not have been as involved in this project without the
kindness of Erik Lindborg.
We are grateful to Bitbucket for providing us with a high quality forge
compatible with Mercurial, free of cost.

\section*{Funding statement}


This project has indirectly benefited from funding from the foundation Simone et
Cino Del Duca de l'Institut de France, the European Research Council (ERC)
under the European Union's Horizon 2020 research and innovation program (grant
agreement No 647018-WATU and Euhit consortium) and the Swedish Research Council
(Vetenskapsr{\aa}det): 2013-5191.
We have also been able to use supercomputers of CIMENT/GRICAD, CINES/GENCI and
the Swedish National Infrastructure for Computing (SNIC).

\section*{Competing interests}

The authors declare that they have no competing interests.


%

\bibliographystyle{agsm}
\bibliography{bib}

\rule{\textwidth}{1pt}

{\bf Copyright Notice} \\
Authors who publish with this journal agree to the following terms: \\

Authors retain copyright and grant the journal right of first publication with
the work simultaneously licensed under a
\href{http://creativecommons.org/licenses/by/3.0/}{Creative Commons Attribution
License} that allows others to share the work with an acknowledgement of the
work's authorship and initial publication in this journal.

Authors are able to enter into separate, additional contractual arrangements
for the non-exclusive distribution of the journal's published version of the
work (e.g., post it to an institutional repository or publish it in a book),
with an acknowledgement of its initial publication in this journal.

By submitting this paper you agree to the terms of this Copyright Notice, which
will apply to this submission if and when it is published by this journal.

\end{document}